\documentclass[a4paper,secnumarabic,amssymb,amsmath,preprint,floats,epsfig]{revtex4}
\usepackage{graphicx}

\renewcommand{\thefootnote}{\fnsymbol{footnote}}

\begin{document}

\preprint{\tighten\vbox{\hbox{\hfil BELLE-CONF-0901}
                        \hbox{\hfil NTLP Preprint 2009-03}
}}


\title{\bf\Large
Measurement of $B \rightarrow D^{(*)} \tau \nu $
using full reconstruction tags}

\affiliation{Budker Institute of Nuclear Physics, Novosibirsk}
\affiliation{Chiba University, Chiba}
\affiliation{University of Cincinnati, Cincinnati, Ohio 45221}
\affiliation{Department of Physics, Fu Jen Catholic University, Taipei}
\affiliation{Justus-Liebig-Universit\"at Gie\ss{}en, Gie\ss{}en}
\affiliation{The Graduate University for Advanced Studies, Hayama}
\affiliation{Gyeongsang National University, Chinju}
\affiliation{Hanyang University, Seoul}
\affiliation{University of Hawaii, Honolulu, Hawaii 96822}
\affiliation{High Energy Accelerator Research Organization (KEK), Tsukuba}
\affiliation{Hiroshima Institute of Technology, Hiroshima}
\affiliation{University of Illinois at Urbana-Champaign, Urbana, Illinois 61801}
\affiliation{India Institute of Technology Guwahati, Guwahati}
\affiliation{Institute of High Energy Physics, Chinese Academy of Sciences, Beijing}
\affiliation{Institute of High Energy Physics, Vienna}
\affiliation{Institute of High Energy Physics, Protvino}
\affiliation{Institute of Mathematical Sciences, Chennai}
\affiliation{INFN - Sezione di Torino, Torino}
\affiliation{Institute for Theoretical and Experimental Physics, Moscow}
\affiliation{J. Stefan Institute, Ljubljana}
\affiliation{Kanagawa University, Yokohama}
\affiliation{Institut f\"ur Experimentelle Kernphysik, Universit\"at Karlsruhe, Karlsruhe}
\affiliation{Korea University, Seoul}
\affiliation{Kyoto University, Kyoto}
\affiliation{Kyungpook National University, Taegu}
\affiliation{\'Ecole Polytechnique F\'ed\'erale de Lausanne (EPFL), Lausanne}
\affiliation{Faculty of Mathematics and Physics, University of Ljubljana, Ljubljana}
\affiliation{University of Maribor, Maribor}
\affiliation{Max-Planck-Institut f\"ur Physik, M\"unchen}
\affiliation{University of Melbourne, School of Physics, Victoria 3010}
\affiliation{Nagoya University, Nagoya}
\affiliation{Nara University of Education, Nara}
\affiliation{Nara Women's University, Nara}
\affiliation{National Central University, Chung-li}
\affiliation{National United University, Miao Li}
\affiliation{Department of Physics, National Taiwan University, Taipei}
\affiliation{H. Niewodniczanski Institute of Nuclear Physics, Krakow}
\affiliation{Nippon Dental University, Niigata}
\affiliation{Niigata University, Niigata}
\affiliation{University of Nova Gorica, Nova Gorica}
\affiliation{Novosibirsk State University, Novosibirsk}
\affiliation{Osaka City University, Osaka}
\affiliation{Osaka University, Osaka}
\affiliation{Panjab University, Chandigarh}
\affiliation{Peking University, Beijing}
\affiliation{Princeton University, Princeton, New Jersey 08544}
\affiliation{RIKEN BNL Research Center, Upton, New York 11973}
\affiliation{Saga University, Saga}
\affiliation{University of Science and Technology of China, Hefei}
\affiliation{Seoul National University, Seoul}
\affiliation{Shinshu University, Nagano}
\affiliation{Sungkyunkwan University, Suwon}
\affiliation{School of Physics, University of Sydney, NSW 2006}
\affiliation{Tata Institute of Fundamental Research, Mumbai}
\affiliation{Excellence Cluster Universe, Technische Universit\"at M\"unchen, Garching}
\affiliation{Toho University, Funabashi}
\affiliation{Tohoku Gakuin University, Tagajo}
\affiliation{Tohoku University, Sendai}
\affiliation{Department of Physics, University of Tokyo, Tokyo}
\affiliation{Tokyo Institute of Technology, Tokyo}
\affiliation{Tokyo Metropolitan University, Tokyo}
\affiliation{Tokyo University of Agriculture and Technology, Tokyo}
\affiliation{Toyama National College of Maritime Technology, Toyama}
\affiliation{IPNAS, Virginia Polytechnic Institute and State University, Blacksburg, Virginia 24061}
\affiliation{Yonsei University, Seoul}
  \author{I.~Adachi}\affiliation{High Energy Accelerator Research Organization (KEK), Tsukuba} 
  \author{H.~Aihara}\affiliation{Department of Physics, University of Tokyo, Tokyo} 
  \author{K.~Arinstein}\affiliation{Budker Institute of Nuclear Physics, Novosibirsk}\affiliation{Novosibirsk State University, Novosibirsk} 
  \author{T.~Aso}\affiliation{Toyama National College of Maritime Technology, Toyama} 
  \author{V.~Aulchenko}\affiliation{Budker Institute of Nuclear Physics, Novosibirsk}\affiliation{Novosibirsk State University, Novosibirsk} 
  \author{T.~Aushev}\affiliation{\'Ecole Polytechnique F\'ed\'erale de Lausanne (EPFL), Lausanne}\affiliation{Institute for Theoretical and Experimental Physics, Moscow} 
  \author{T.~Aziz}\affiliation{Tata Institute of Fundamental Research, Mumbai} 
  \author{S.~Bahinipati}\affiliation{University of Cincinnati, Cincinnati, Ohio 45221} 
  \author{A.~M.~Bakich}\affiliation{School of Physics, University of Sydney, NSW 2006} 
  \author{V.~Balagura}\affiliation{Institute for Theoretical and Experimental Physics, Moscow} 
% ...
\collaboration{The Belle Collaboration}

\vspace {1.0cm}

\begin{abstract}

We present measurements of $B \rightarrow D^{*} \tau \nu $ and
$B \rightarrow D \tau \nu$ decays using 604.5 fb$^{-1}$ of data collected 
at the $\Upsilon(4S)$ resonance with the Belle detector at the KEKB 
asymmetric-energy $e^+e^-$ collider.
Events are tagged by fully reconstructing one of the $B$ mesons in hadronic 
modes.
We obtain 
${\cal B}\,(B^+ \rightarrow \overline{D}{}^0 \tau^+ \nu )$ 
$= (1.51~^{+0.41}_{-0.39}~^{+0.24}_{-0.19}~\pm 0.15) \%$,
${\cal B}\,(B^+ \rightarrow \overline{D}{}^{*0} \tau^+ \nu )$ 
$= (3.04~^{+0.69}_{-0.66}~^{+0.40}_{-0.47}~\pm 0.22) \%$,
${\cal B}\,(B^0 \rightarrow D^- \tau^+ \nu )$ 
$= (1.01~^{+0.46}_{-0.41}~^{+0.13}_{-0.11}~\pm 0.10)\%$,
${\cal B}\,(B^0 \rightarrow D^{*-} \tau^+ \nu )$ 
$= (2.56~^{+0.75}_{-0.66}~^{+0.31}_{-0.22}~\pm 0.10)\%$
where the first error is statistical, the second is 
systematic, and the third is due to the uncertainty in the branching fraction
for the normalization mode.

\end{abstract}

\maketitle

%
%
%

{\renewcommand{\thefootnote}{\fnsymbol{footnote}}

\normalsize

\setcounter{footnote}{0}

\clearpage

\normalsize



\setcounter{page}{1}

\clearpage

\section{Introduction}
\label{sec:intro}
The semileptonic $B$ decay to $\tau$ channel, $B \to D^{(*)} \tau \nu$,
is a sensitive probe of extensions to the Standard Model (SM).
In the SM, it occurs via an external $W$ emission diagram with predicted 
branching fractions of $(0.69 \pm 0.04)$\% and $(1.41 \pm 0.07)$\% 
for the $B^0 \to D^{-} \tau^+ \nu_{\tau}$ and 
$B^0 \to D^{*-}\tau^+\nu_{\tau}$ modes, respectively~\cite{Chen06}.
On the other hand, in extensions of the SM, such as the Two Higgs Doublet Models 
(2HDM) and the Minimal Supersymmetric Standard Model (MSSM), a charged Higgs 
boson ($H^{\pm}$) may contribute to the decay amplitude at tree level, and 
the branching fraction may be modified significantly~\cite{Grzad92, Tanaka95,
Kiers97, Itoh05, Nierste07}.
Both $B \to D^{(*)} \tau \nu$ and the purely leptonic decay
$B^+ \to \tau^+ \nu_{\tau}$ have similar sensitivity to $H^{\pm}$ bosons
with different theoretical systematics; the former suffers from uncertainty 
in the form factor, while the latter requires knowledge of the $B$ decay
constant $f_{B}$.
Therefore, the two decays provide complementary approaches to searching for 
$H^{\pm}$ signatures in $B$ decays.

Experimentally, measurements of the $B \to D^{(*)} \tau \nu$ decays are
challenging because at least two neutrinos are present in the final state.
The Belle collaboration has previously reported the first observation of the 
decay $B^0 \to D^{*-} \tau^+ \nu$, by inclusively reconstructing the 
accompanying $B$ via a 4-vector sum of all the charged and neutral tracks other 
than the $D^*$ and $\tau$ daughter track candidates.
The reported branching fraction is 
${\cal B}\,(B^0 \to D^{*-} \tau^+ \nu) = 
(2.02_{-0.37}^{+0.40}({\rm stat}) \pm 0.37 ({\rm syst}))$\%~\cite{Belle_Dsttaunu}.
In this paper, we present new measurements of $B^0 \to D^{(*)-} \tau^+ \nu$ 
and $B^+ \to \overline{D}{}^{(*)0} \tau^+ \nu$ decays.
Here we fully reconstruct one of the $B$ mesons in the event, referred to 
hereafter as the tag side ($B_{\rm tag}$), and compare properties of 
the remaining particles, referred to as the signal side ($B_{\rm sig}$), 
to those expected for signal and background.
The method allows us to strongly suppress the combinatorial backgrounds,
and correctly calculate the missing mass which discriminates the signal
from $B \to D^{(*)} \ell \nu$ background.
Using a similar technique, the BaBar collaboration has reported the
branching fractions, 
${\cal B}(B^0 \to D^- \tau^+ \nu) = (0.86 \pm 0.24 \pm 0.11 \pm 0.06)$\% and  
${\cal B}(B^0 \to D^{*-} \tau^+ \nu) = (1.62 \pm 0.31 \pm 0.10 \pm 0.05)$\%,
where the third uncertainty is from the normalization mode.
They also measured distributions of the lepton momentum and the squared
momentum transfer~\cite{BaBar_Dtaunu}.

In order to avoid experimental bias, the signal region in data is blinded
until the event selection criteria are finalized. 
Inclusion of the charge conjugate decays is implied throughout this paper.

\section{Experiment and Data Set}
\label{sec:data_exp}
The analysis is based on the data recorded with the Belle detector at the 
KEKB $e^+e^-$ asymmetric-energy collider operating at the center-of-mass 
(c.m.) energy of the $\Upsilon(4S)$ resonance.
KEKB consists of a low energy ring (LER) of 3.5\,GeV positrons and a high
energy ring (HER) of 8\,GeV electrons~\cite{KEKB}.
The $\Upsilon(4S)$ data set used in this analysis corresponds to an
integrated luminosity of $605$ fb$^{-1}$ and contains 
$657 \times 10^6$ $B \overline{B}$ events.

The Belle detector is a large-solid-angle magnetic spectrometer
that consists of a silicon vertex detector (SVD),
a 50-layer central drift chamber (CDC), 
an array of aerogel threshold Cherenkov counters (ACC),
a barrel-like arrangement of time-of-flight scintillation counters (TOF),
and an electromagnetic calorimeter (ECL) comprised of CsI(Tl) crystals
located inside a super-conducting solenoid coil 
that provides a 1.5~T magnetic field.  
Muons and $K_L^0$ mesons are identified by arrays of resistive plate counters
interspersed in the iron yoke (KLM).
The detector is described in detail elsewhere~\cite{BELLE}.

A detailed Monte Carlo (MC) simulation, based on the GEANT package~\cite{GEANT}, 
is used to estimate the signal detection efficiency and to study the 
background.
Large samples of the signal decays are generated with the EVTGEN package~\cite{Evtgen}
using the ISGW~II form factor model~\cite{ISGW2}.
To model the $B\overline{B}$ and $q\overline{q} (q = u, d, s, c)$ backgrounds, large 
generic $B\overline{B}$ and $q\overline{q}$ MC samples, corresponding to 
about twice the integrated luminosity of data are used.
To further increase the $B\overline{B}$ MC statistics, we also use special 
$B\overline{B}$ MC samples, corresponding to $1.5 \times 10^{10}$
$B^0\overline{B}{}^0$ and $B^+B^-$ pairs, where events are filtered based on 
event generator information before running the time consuming GEANT 
detector simulation.

\section{Event Reconstruction and Selection}
\label{sec:recon}

\subsection{Tag-side Reconstruction}
Charged particle tracks are reconstructed from hits in the SVD and CDC. 
They are required to satisfy track quality cuts based on their impact 
parameters relative to the measured profile of the interaction point 
(IP) of the two beams. 
Charged kaons are identified by combining information on ionization loss 
($dE/dx$) in the CDC, Cherenkov light yields in the ACC and 
time-of-flight measured by the TOF system.
For the nominal requirement, the kaon identification efficiency is 
approximately $88\%$ and the probability of misidentifying a pion as 
a kaon is about $8\%$.
Hadron tracks that are not identified as kaons are treated as pions.
Tracks satisfying the lepton identification criteria, as described below, 
are removed from consideration.
Electron identification is based on a combination of $dE/dx$ in CDC,
the response of the ACC, shower shape in the ECL, 
position matching between ECL clusters and the track, and the ratio of the
energy deposited in the ECL to the momentum measured by the tracking system.
Muon candidates are selected using range of tracks measured in KLM
and the deviation of hits from the extrapolated track trajectories.
The lepton identification efficiencies are estimated to be about 90\% 
for both electrons and muons in the momentum region above 1.2\,GeV/$c$. 
The hadron misidentification rate is measured using reconstructed 
$K_S^0 \to \pi^+ \pi^-$ decays and found to be less than 0.2\% for electrons 
and 1.5\% for muons in the same momentum region.
$K_S^0$ mesons are reconstructed using pairs of charged tracks that
have an invariant mass within $\pm 30$\,MeV/$c^2$ of the known $K_S^0$ 
mass and a well reconstructed vertex that is displaced from the IP.
Candidate $\gamma$'s are required to have a minimum energy deposit
$E_{\gamma} \geq 50$\,MeV.
Candidate $\pi^0$ mesons are reconstructed using $\gamma\gamma$ pairs 
with invariant masses between 117 and 150\,MeV/$c^2$.
For slow $\pi^0$'s used in $D^{*}$ reconstruction, the minimum
$\gamma$ energy requirement is lowered to 30\,MeV.

$B_{\rm tag}$ candidates are reconstructed in the following decay modes: 
$B^{+} \rightarrow \overline{D}{}^{(*)0} \pi^{+}$, 
$\overline{D}{}^{(*)0}\rho^{+}$, 
$\overline{D}{}^{(*)0}a_{1}^{+}$, 
$\overline{D}{}^{(*)0}D_{s}^{(*)+}$, and
$B^{0} \rightarrow D^{(*)-} \pi^{+}$, 
$D^{(*)-}\rho^{+}$, 
$D^{(*)-}a_{1}^{+}$, 
$D^{(*)-}D_{s}^{(*)+}$.
Candidate $\rho^+$ and $\rho^0$ mesons are reconstructed in the 
$\pi^+\pi^0$ and $\pi^+\pi^-$ decay modes, by requiring their invariant 
masses to be within $\pm 225$\,MeV/$c^2$ of the nominal $\rho$ mass.
We then select $a_1^+$ candidates by combining a $\rho^0$ candidate
and a pion with invariant mass between 0.7 and 1.6\,GeV/$c^2$
and require that the charged tracks form a good vertex. 
$D$ meson candidates are reconstructed in the following decay modes: 
$\overline{D}{}^{0}\rightarrow K^{+}\pi^{-}$, $K^{+}\pi^{-}\pi^{0}$,
$K^{+}\pi^{-}\pi^{+}\pi^{-}$, $K_{S}^{0}\pi^{0}$, 
$K_{S}^{0}\pi^{-}\pi^{+}$, $K_{S}^{0}\pi^{-}\pi^{+}\pi^{0}$, 
$K^{-}K^{+}$, and
$D^- \rightarrow K^{+}\pi^{-}\pi^{-}$, $K^{+}\pi^{-}\pi^{-}\pi^{0}$,
$K_{S}^{0}\pi^{-}$, $K_{S}^{0}\pi^{-}\pi^{0}$, 
$K_{S}^{0}\pi^{+}\pi^{-}\pi^{-}$, $K^+K^-\pi^-$,
$D_{s}^{+}\rightarrow K_{S}^{0}K^{+}$, $K^{+}K^{-}\pi^{+}$.
The $D$ candidates are required to have an invariant mass $M_D$ within
$4-5 \sigma$ ($\sigma$ is the mass resolution) 
of the nominal $D$ mass value depending on the mode.
$D^{*}$ mesons are reconstructed as
$D^{*+} \rightarrow D^{0}\pi^{+}$, $D^{+}\pi^{0}$,
$D^{*0} \rightarrow D^{0}\pi^{0}$, $D^{0}\gamma$, and
$D_{s}^{*+} \rightarrow D_{s}^{+}\gamma$.
$D^*$ candidates from modes that include a pion are required to have a
mass difference $\Delta M = M_{D\pi} - M_{D}$ within $\pm 5$\,MeV/$c^2$
of its nominal value.
For decays with a photon, we require that the mass difference
$\Delta M = M_{D\gamma} - M_{D}$ be within $\pm 20$\,MeV/$c^2$ of the
nominal value. 

The selection of $B_{\rm tag}$ candidates is based on the 
beam-constrained mass $M_{\rm bc}\equiv\sqrt{E_{\rm beam}^{2} - p_{B}^{2}}$
and the energy difference $\Delta E\equiv E_{B} - E_{\rm beam}$.
Here, $E_{B}$ and $p_{B}$ are the reconstructed energy and momentum
of the $B_{\rm tag}$ candidate in the $e^+e^-$ c.m. system,
and $E_{\rm beam}$ is the beam energy in the c.m. frame.
The background from jet-like continuum events ($e^+e^- \to q\overline{q}, q=u,d,s,c$)
is suppressed on the basis
of event topology: we require the normalized second Fox-Wolfram moment
($R_2$)~\cite{R2} to be smaller than 0.5, and 
$|\cos\theta_{\rm th}|<0.8$, where $\theta_{\rm th}$ is the angle between
the thrust axis of the $B$ candidate and that of the remaining tracks in 
the event.
The latter requirement is not applied to 
$B^{+} \rightarrow \overline{D}{}^{0}\pi^+$,
$\overline{D}{}^{*0}(\rightarrow \overline{D}{}^{0}\pi^0)\pi^{+}$ and
$B^{0} \rightarrow D^{*-} (\rightarrow \overline{D}{}^{0}\pi^-)\pi^+$
decays, where the continuum background is small.
For the $B_{\rm tag}$ candidate, we require 
$5.27~\mbox{GeV}/c^{2}<M_{\rm bc}<5.29~\mbox{GeV}/c^{2}$ and 
$-80~\mbox{MeV} <\Delta E< 60~\mbox{MeV}$.
If an event has multiple $B_{\rm tag}$ candidates, we choose the one having
the smallest $\chi^{2}$ based on deviations from the nominal values of 
$\Delta E$, the $D$ candidate mass, and the $D^{*} - D$ mass difference if
applicable.
The number of $B^+$ and $B^0$ candidates in the selected region are
$1.75 \times 10^6$ and $1.18 \times 10^6$, respectively.
By fitting the distribution to the sum of an empirical parameterization of 
the background shape~\cite{Albrecht} and a signal shape~\cite{Bloom:1983pc}, 
we estimate that in the selected region there are 
$(10.11 \pm 0.03) \times 10^5$~(with purity=0.58) $B^+$ and
$( 6.05 \pm 0.03) \times 10^5$~(with purity=0.51) $B^0$ events,
respectively.

\subsection{Signal-side Reconstruction}
In the events where a $B_{\rm tag}$ is reconstructed, we search for decays
of $B_{\rm sig}$ into a $D^{(*)}$, $\tau$ and a neutrino. 
In the present analysis, the $\tau$ lepton is identified in the leptonic 
decay modes, $\mu^{-}\overline{\nu} \nu$ and $e^{-}\overline{\nu} \nu$.
We require that the charge/flavor of the $\tau$ daughter particles and 
the $D$ meson are consistent with the $B_{\rm sig}$ flavor, opposite
to the $B_{\rm tag}$ flavor.
The loss of signal due to $B^0 - \overline{B^0}$ mixing is estimated by
the MC simulation.  

The procedures to reconstruct charged particles ($e^{\pm}, \mu^{\pm}, 
\pi^{\pm}, K^{\pm}$) and neutral particles ($\pi^0, K_S^0$) for the 
signal side are the same as those used for the tag side.
For $\gamma$ candidates, 
we require a minimum energy threshold of 50\,MeV for the barrel, and 100
(150)\,MeV for the forward (backward) end-cap ECL.
A higher threshold is used for the endcap ECL, where the effect of beam 
background is more severe.
We also require that the lepton momentum in the laboratory frame
exceeds 0.6\,GeV/$c$ to ensure good lepton identification efficiency.

The decay modes used for $D$ reconstruction are slightly different from 
those used for the tagging side: 
$\overline{D}{}^{0}\rightarrow K^{+}\pi^{-}$, $K^{+}\pi^{-}\pi^{0}$,
$K^{+}\pi^{-}\pi^{+}\pi^{-}$, $K^{+}\pi^{-}\pi^{+}\pi^{-}\pi^0$, 
$K_{S}^{0}\pi^{0}$, $K_{S}^{0}\pi^{-}\pi^{+}$, 
$K_{S}^{0}\pi^{-}\pi^{+}\pi^{0}$, and
$D^- \rightarrow K^{+}\pi^{-}\pi^{-}$, $K^{+}\pi^{-}\pi^{-}\pi^{0}$,
$K_{S}^{0}\pi^{-}$.
The $D$ candidates are required to have an invariant mass $M_D$ within
$5 \sigma$ of the nominal $D$ mass value.
$D^{*}$ mesons are reconstructed using the same decay modes as on the
tagging side:
$D^{*+} \rightarrow D^{0}\pi^{+}$, $D^{+}\pi^{0}$, and
$D^{*0} \rightarrow D^{0}\pi^{0}$, $D^{0}\gamma$.
$D^*$ candidates are required to have a mass difference 
$\Delta M = M_{D\pi(\gamma)} - M_{D}$ within $5 \sigma$ of the nominal value.

For signal selection, we use the following variables that characterize the 
signal decay: the missing mass squared in the event ($M_{\rm miss}^2$), the momentum
(in the c.m. system) of the $\tau$ daughter leptons ($P_\ell^*$), and the extra 
energy in the ECL ($E_{\rm extra}^{\rm ECL}$).
The missing mass squared is calculated as 
$ M_{\rm miss}^2 = (E_{B_{\rm tag}}-E_D-E_{\tau\rightarrow X})^2 - 
(-\vec{P}_{B_{tag}}-\vec{P}_{D^{(*)}} - \vec{P}_{\tau\rightarrow X})^2 $,
using the energy and momenta of the $B_{\rm tag}$, the $D^{(*)}$ candidate and 
the lepton from the $\tau$ decay.
The signal decay is characterized by large $M_{\rm miss}^2$ due to the presence of 
more than two neutrinos in the final state.
The lepton momenta ($P_\ell^*$) distribute 
lower than those from primary $B$ decays.
The extra energy in the ECL ($E_{\rm extra}^{\rm ECL}$) is the sum of
the energies of photons that are not associated with either the $B_{\rm tag}$ 
or the $B_{\rm sig}$ reconstruction.
ECL clusters with energies greater than 50\,MeV in the barrel, and 100
(150)\,MeV in the forward (backward) end-cap ECL are used to calculate
$E_{\rm extra}^{\rm ECL}$.
For signal events, $E_{\rm extra}^{\rm ECL}$ must be either zero or a 
small value arising from beam background hits, therefore, signal events 
peak at low $E_{\rm extra}^{\rm ECL}$.
On the other hand, background events are distributed toward higher 
$E_{\rm extra}^{\rm ECL}$ due to the contribution from additional neutral 
clusters.
We also require that the event has no extra charged tracks and no $\pi^0$ 
candidate other than those from the signal decay and those used in the 
$B_{\rm tag}$ reconstruction.
Table~\ref{tbl:selection} summarizes the cuts to define the signal region.
The cuts are optimized by maximizing the figure of merit (F.O.M.), defined 
as ${\rm F.O.M.} = N_S/\sqrt{N_S+N_B}$, where $N_S(N_B)$ are the 
number of signal (total background) events in the signal region, assuming
the SM branching fractions for the $D \tau \nu$ and the $D^* \tau \nu$ modes.

\begin{table}[htbp]
 \begin{center}
  \begin{tabular}{lll}
   \hline\hline
   Cut variable~~
   & $B \to \overline{D}{}^0(D^-) \tau^+ \nu$~~
   & $B \to \overline{D}{}^{*0}(D^{*-}) \tau^+ \nu$ \\
   \hline 
   Number of extra tracks  & $=0$                    &$=0$                    \\
   Number of extra $\pi^0$ & $=0$                    &$=0$                    \\
   $P_\ell^*$           &$\leq 1.2$\,GeV/$c$       &$\leq 1.2$\,GeV/$c$       \\ 
   $M_{\rm miss}^2$                     &$\geq 2.0$\,GeV$^2$/$c^4$ &$\geq 1.6$\,GeV$^2$/$c^4$ \\ 
   $E_{\rm extra}^{\rm ECL}$  &$\leq 0.2$\,GeV           &$\leq 0.2$\,GeV           \\
   \hline\hline
  \end{tabular}
  \caption{Summary of the signal selection criteria.}
  \label{tbl:selection}
 \end{center}
\end{table}

\subsection{Signal Detection Efficiency and Expected Background}
Table~\ref{tbl:efficiency} lists the signal detection efficiencies, which are
estimated from signal MC simulation, with the selection criteria 
shown in Table~\ref{tbl:selection}.
Taking account of the cross talks between $B \to D \tau \nu$ and 
$B \to D^{*} \tau \nu$ modes, the signal detection efficiency 
($\epsilon_{ij}$) is defined as,
\begin{equation}
N_{ij} = \epsilon_{ij} \cdot {\cal B}{}_j \cdot N_{\rm tag}~~~~~,  
\label{eqn:efficiency}
\end{equation}
where $N_{ij}$ represents the yield of the generated $j$-th mode 
reconstructed in the $i$-th mode.
${\cal B}{}_j$ is the branching fraction of the $j$-th mode including 
the sub-decay ($\tau$ and $D^{(*)}$) branching fractions.
$N_{\rm tag}$ is the number of $B$ events fully reconstructed on the tagging
side.
Table~\ref{tbl:efficiency} also shows, in parentheses, the efficiencies without 
cuts on $M_{\rm miss}^2$ and $E_{\rm extra}^{\rm ECL}$.
These are the two variables used to extract the signal yields.

\begin{table}[htbp]
 \begin{center}

  \begin{tabular}{l|ll}
   \hline\hline
   Recon'd mode & \multicolumn{2}{c}{Generated modes} \\
& $\overline{D}{}^0 \tau^+ \nu$ & $\overline{D}{}^{*0} \tau^+ \nu$ \\
   \hline
$\overline{D}{}^0 \tau^+ \nu$    
& $2.55 \pm 0.05 $ ($4.87 \pm 0.08 $)~ & $0.90 \pm 0.05 $ ($1.75 \pm 0.07$) \\
$\overline{D}{}^{*0} \tau^+ \nu$ 
& $0.34 \pm 0.01 $ ($1.33 \pm 0.02 $)~ & $1.08 \pm 0.03 $ ($2.11 \pm 0.04$) \\
   \hline\hline
  \end{tabular}

  \vspace{0.5cm}

  \begin{tabular}{l|ll}
   \hline\hline
   Recon'd mode & \multicolumn{2}{c}{Generated modes} \\
                    & $D^-    \tau^+ \nu$ & $D^{*-} \tau^+ \nu$ \\
   \hline
$D^-    \tau^+ \nu$ 
& $3.21 \pm 0.06 $ ($6.86 \pm 0.09 $)~ & $0.23 \pm 0.03$ ($0.55 \pm 0.03 $) \\
$D^{*-} \tau^+ \nu$ 
& $0.11 \pm 0.01 $ ($0.27 \pm 0.01 $)~ & $0.80 \pm 0.02$ ($1.54 \pm 0.03 $) \\
   \hline\hline
  \end{tabular}

  \caption{Signal detection efficiency (\%) matrix for $B^{+}$ (top) and
$B^0$ (bottom) modes. The values in parenthesis are the efficiencies without 
cuts on $M_{\rm miss}^2$ and $E_{\rm extra}^{\rm ECL}$.}
  \label{tbl:efficiency}
 \end{center}
\end{table}
  
According to the MC simulation, the expected number of signal (background)
events in the signal region is 
19(48) for $B^+ \to \overline{D}{}^0 \tau^+ \nu$, 
7(13)  for $B^0 \to D^- \tau^+ \nu$,  
18(25) for $B^+ \to \overline{D}{}^{*0} \tau^+ \nu$, and
7(6)   for $B^0 \to D^{*-} \tau^+ \nu$.
The major background sources are semileptonic $B$ decays, $B \to D \ell \nu$, 
$D^* \ell \nu$ and $D^{**} \ell \nu$ (60-70\% depending on the decay 
mode).
The remaining background comes from hadronic $B$ decays including a $D$ meson
in the final state. 
Background from $q\overline{q}$ processes are found to be small (less than
one event).
As shown in Table~\ref{tbl:efficiency}, the cross talk between $B \to D \tau \nu$
and $B \to D^{*} \tau \nu$ arises, when a pion or a photon is missed in 
the reconstruction of $D^*$, or when a random photon is combined with a $D$
to form a fake $D^*$. The cross-feed to other $B^0$ or $B^+$ tag samples is 
negligibly small.

\section{Calibration using the $B \to D^{(*)} \ell \nu$ Sample}
\label{sec:controlsample}

We use $B \to D^{(*)} \ell \nu$ ($\ell = e/\mu$) decays as control samples 
to calibrate the background MC simulation and to verify the 
$E_{\rm extra}^{\rm ECL}$ simulation.
We also use these decays to normalize the extracted signal yields.
We select $B \to D^{(*)} \ell \nu$ decays using the same selection 
requirements as $B \to D^{(*)} \tau \nu$, but without the cut on the 
momentum of the $\tau$ daughter lepton and 
with $|M_{\rm mis}^2|<1 {\rm\, GeV}^2/c^4$ and $E_{\rm extra}^{\rm ECL}<1.2$\,GeV.
The four calibration decay modes: 
$B^+ \to \overline{D}{}^0 \ell^+ \nu$, 
$B^+ \to \overline{D}{}^{*0} \ell^+ \nu$, 
$B^0 \to D^-   \ell^+ \nu$, and
$B^0 \to D^{*} \ell^+ \nu$,
peak around zero in the missing mass distributions,
as shown in Figure~\ref{fig:dlnu_mm2}.

The yields of the calibration modes are extracted by fitting the distributions 
with expected shapes based on MC simulation for the signal and the background.
The major background in each distribution arises from other semileptonic 
decays, where a pion or a photon is missed (i.e. $B \to D^* \ell \nu$ is 
reconstructed as $B \to D \ell \nu$ if the soft $\pi^0$ or $\gamma$ from 
the $D^*$ is missed), or a random photon is used in $D^{*0}$ reconstruction 
(i.e. $B \to D \ell \nu$ misreconstructed as $B \to D^* \ell \nu$).
Here the two distributions for $B^+$ and $B^0$ candidates are fitted simultaneously.
Table~\ref{tbl:calibresults} lists the yields extracted for each calibration 
decay mode, which include the yields detected as cross talk; for example,
the yield of $\overline{D}{}^0 \ell^+ \nu$ is the sum of $B^+ \to 
\overline{D}{}^0 \ell^+ \nu$ decays measured in the $\overline{D}{}^0 \ell^+ 
\nu$ and $\overline{D}{}^{*0} \ell^+ \nu$ distributions.  
When we compare the extracted yields with expected yields from the MC 
simulation, the ratio of the measured to the expected yields 
($R_{\rm corr}$) are found to be $0.75$ -- $ 0.84$, depending on the mode. 
The ratios are used as scale factors to correct the normalization in the
MC simulation for $B \to D^{(*)} \ell \nu$ semileptonic decays, which are 
the major backgrounds in the $B \to D^{(*)} \tau \nu$ detection.

\begin{table}[htbp]
 \begin{center}
  \begin{tabular}{lllll}
   \hline\hline
   & 
   $\overline{D}{}^0 \ell^+ \nu$ &
   $\overline{D}{}^{*0} \ell^+ \nu$ &
   $\overline{D}{}^- \ell^+ \nu$ &
   $\overline{D}{}^{*-} \ell^+ \nu$  \\
   \hline
   Yield &
   1156 $\pm$ 44 & 2152 $\pm$ 76 &
    338 $\pm$ 21 &  769 $\pm$ 35 \\
   Efficiency (\%) &
   8.97 $\pm$ 0.05 & 6.86 $\pm$ 0.08    &
   11.3 $\pm$ 0.12 & 5.43 $\pm$ 0.04   \\
   \hline\hline

  \end{tabular}
  \caption{Yields and efficiencies of the calibration modes.}
  \label{tbl:calibresults}
 \end{center}
\end{table}

\begin{figure}[htbp]
\begin{center}
 \includegraphics[width=0.9\textwidth,keepaspectratio,clip]{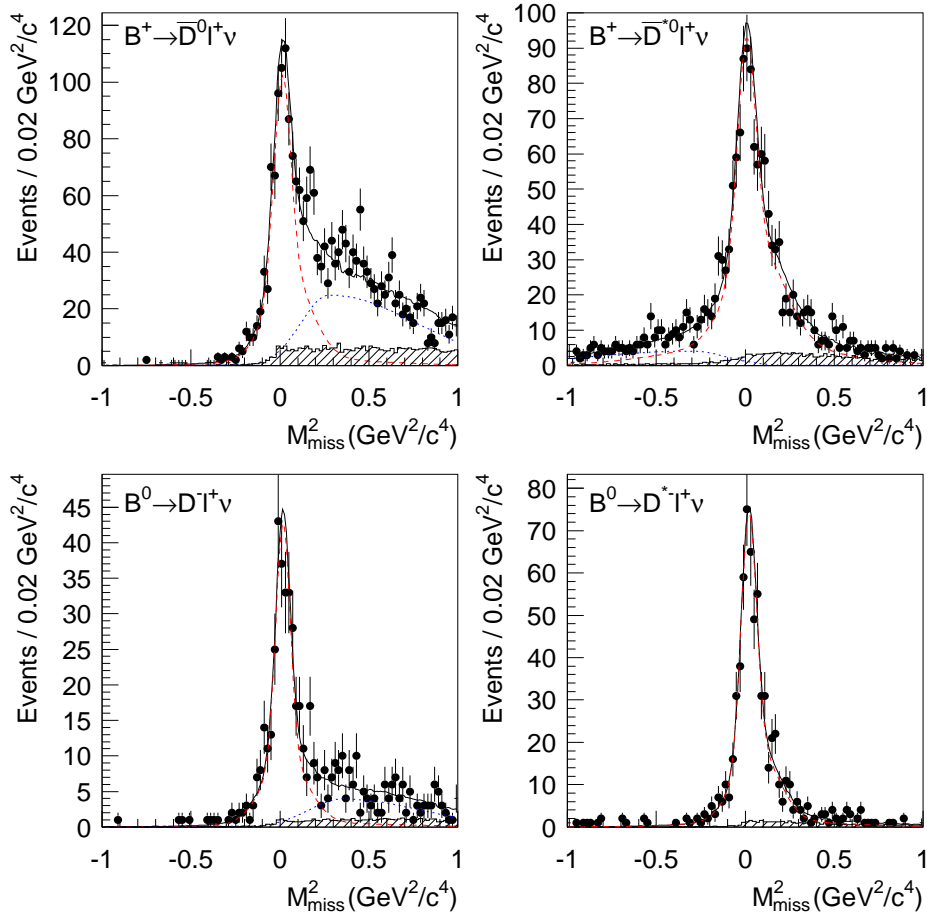}
\caption{
Distribution of missing mass squared ($M_{\rm mis}^2$) for 
$B^+ \to \overline{D}{}^0 \ell^+ \nu$ (top-left), 
$B^+ \to \overline{D}{}^{*0} \ell^+ \nu$ (top-right), 
$B^0 \to D^-   \ell^+ \nu$ (bottom-left), and
$B^0 \to D^{*-} \ell^+ \nu$ (bottom-right).
Data are plotted as points with error bars, the results of the fit (solid line) along with
the signal (dashed red line), other semileptonic decays (dotted blue line) and misidentified
hadronic events (hatched histogram) components are also shown. 
}
\label{fig:dlnu_mm2}
\end{center}
\end{figure}

Figure~\ref{fig:dlnu_ecl} compares the $E_{\rm extra}^{\rm ECL}$ distribution
for the control samples in data and the MC simulation after the correction.
The agreement between the data and the MC simulation is satisfactory, and
valid the $E_{\rm extra}^{\rm ECL}$ simulation.
We also confirm that the number of events found in the sideband of 
the $(M_{\rm miss}^2, E_{\rm extra}^{\rm ECL})$ signal region is consistent
within statistics for the data and the scaled MC simulation.
Here the sideband is defined by $E_{\rm extra}^{\rm ECL}>0.4$\,GeV, 
and $M_{\rm miss}^2 < 1.0$\,GeV$^2$/c$^4$, for all four signal modes. 

\begin{figure}[htbp]
\begin{center}
 \includegraphics[width=0.9\textwidth,keepaspectratio,clip]{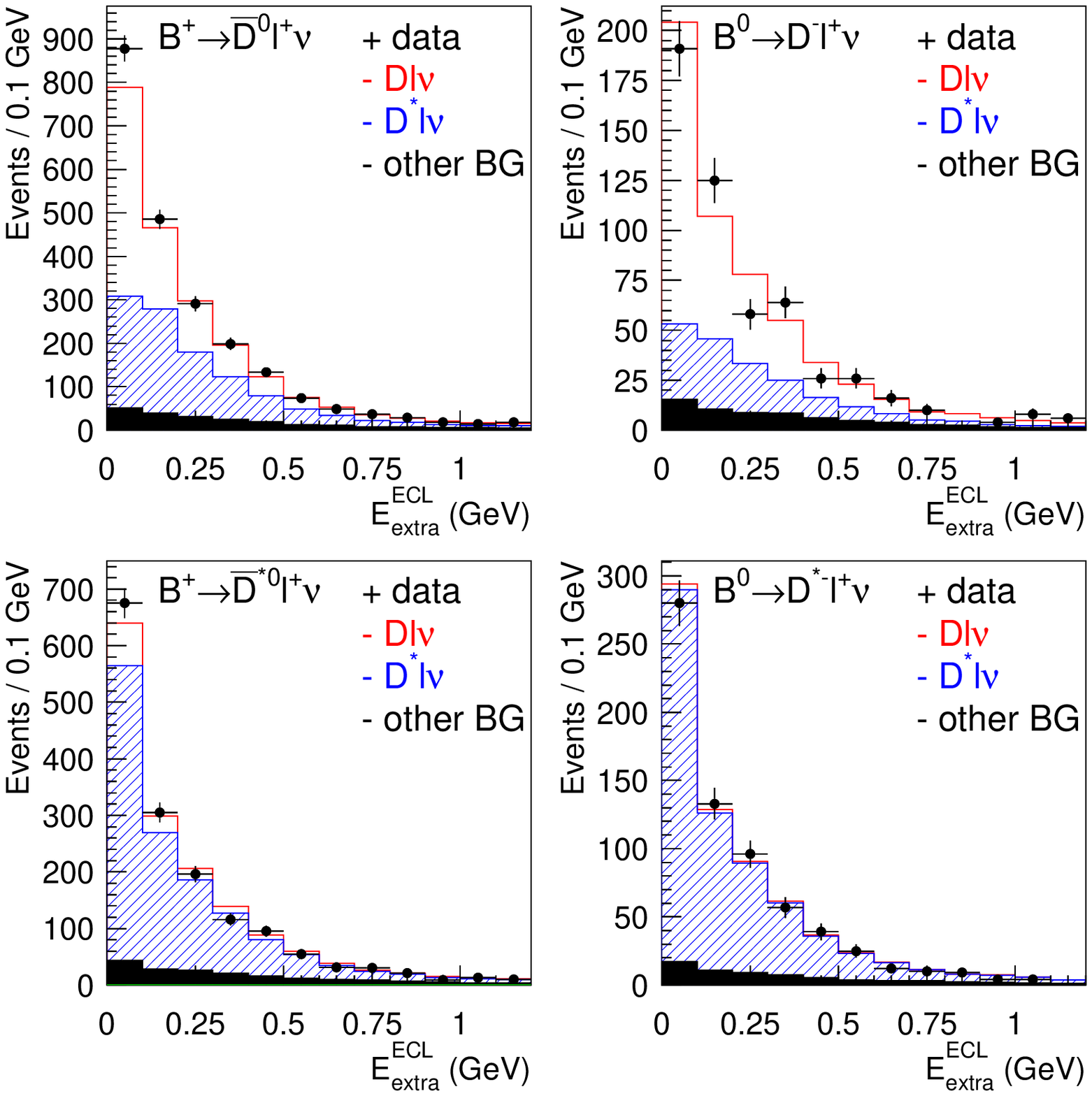}
\caption{
Comparison of $E_{\rm extra}^{\rm ECL}$ distributions for the control samples 
in data and the MC simulation.
}
\label{fig:dlnu_ecl}
\end{center}
\end{figure}

\section{Signal Extraction}
\label{sec:fitting}

The $B \to D \tau \nu$ and $B \to D^{*} \tau \nu$ signal yields are extracted
using unbinned extended maximum likelihood fits to the two-dimensional
$(M_{\rm miss}^2, E_{\rm extra}^{\rm ECL})$ distributions obtained after the 
selection of the signal decays.
The fit components are two signal modes; $B \to D \tau \nu$ and 
$B \to D^{*} \tau \nu$, and the backgrounds from $B \to D \ell \nu$,
$B \to D^* \ell \nu$ and other processes.  
The likelihood is constructed as,
\begin{equation}
L = \frac{e^{-\sum_j N_j}}{N!}\prod_{i=1}^{N} F(M_{\rm miss}^2, E_{\rm extra}^{\rm ECL})
~~~~~,
\end{equation}
where
\begin{equation}
\begin{array}{lll}
F(M_{\rm miss}^2,E_{\rm extra}^{\rm ECL}) & = & N_{D\tau\nu}f_{D\tau\nu}(M_{\rm miss}^2,E_{\rm extra}^{\rm ECL}) + 
N_{D^*\tau\nu}f_{D^*\tau\nu}(M_{\rm miss}^2,E_{\rm extra}^{\rm ECL}) \\
& + & N_{D\ell\nu}f_{D\ell\nu}(M_{\rm miss}^2,E_{\rm extra}^{\rm ECL}) 
+ N_{D^*\ell\nu}f_{D^*\ell\nu}(M_{\rm miss}^2,E_{\rm extra}^{\rm ECL}) \\
& + & N_{\rm other}f_{\rm other}(M_{\rm miss}^2,E_{\rm extra}^{\rm ECL})~~~~~. 
\end{array}
\end{equation}
Here $N_j$ and $f_j(M_{\rm miss}^2,E_{\rm extra}^{\rm ECL})$ represent the number of events and the
two-dimensional probability density function (PDF) as a function
of $M_{\rm miss}^2$ and $E_{\rm extra}^{ECL}$, respectively, for 
process $j$.
In the fit to the $B^0 \to D^{*-} \tau^+ \nu$ distribution, 
the $D \tau \nu$ cross-feed ($f_{D \tau \nu}$) and $D \ell \nu$ background 
($f_{D \ell \nu}$) are not included, because their contribution are found 
to be small according to the MC simulation.
The fit region is defined by 
$(-2 \,{\rm GeV}^2/c^4 < M_{\rm miss}^2 < 8 \,{\rm GeV}^2/c^4, 
0 \,{\rm GeV} < E_{\rm extra}^{\rm ECL} < 1.2 \,{\rm GeV})$
for all four signal modes.

The two-dimensional PDF's for $D^{(*)} \tau \nu$ and $D^{(*)} \ell \nu$  
processes are obtained by taking the product of a one-dimensional PDF for 
each variable, as the correlations between $M_{\rm miss}^2$ and 
$E_{\rm extra}^{\rm ECL}$ 
are found to be small in the MC simulation.
The one-dimensional PDF's for $M_{\rm miss}^2$ are modeled by asymmetric 
Gaussian or double Gaussian distributions, whereas the PDF's for 
$E_{\rm extra}^{\rm ECL}$ are histograms obtained 
from the MC simulation.
The PDF for other background processes ($f_{\rm other}$) uses the two-dimensional 
histograms obtained from MC simulation, 
since correlations between the two variables are significant for these 
background processes, which mainly come from hadronic $B$ decays.

We fit the distributions for $B^0$ and $B^+$ tags separately.
The cross talk between the two tags is found to be small according 
to the MC simulation. 
For each $B^0$ and $B^+$ tag, we then fit simultaneously the two 
distributions for the $D \tau \nu$ and $D^* \tau \nu$ components.
The ratio of the number of events found in the two distributions
is constrained according to the efficiency matrix shown in Table
~\ref{tbl:efficiency}.  

The above procedure to extract the signal yields has been tested by 
``toy MC experiments'': in each experiment, the number of events in 
each $(M_{\rm miss}^2, E_{\rm extra}^{\rm ECL})$ bin is generated according to 
Poisson statistics, with the mean ($\mu$) fixed to the number 
of events found in the MC simulation, including the
$B \to D^{(*)} \tau \nu$ signals with the SM branching fraction.
The distributions are then fit with the procedure described in the  
previous subsection.
We generate 500 experiments, and we confirm 
that the means of the extracted yields are consistent with
the input $\mu$ values.

The signal extraction procedure has also been checked by performing 
a fit to sample distributions from generic MC events, which are the 
sum of the generic 
$B\overline{B}$ and $q\overline{q}$ processes,
where semileptonic $B$ to $\tau$ decays,
$B \to D \tau \nu$, $D^* \tau \nu$ and $D^{**} \tau \nu$,
are removed from the $B\overline{B}$ samples.
For all four signal decays, the signal yields obtained are consistent 
with zero within the statistical uncertainty.

\section{Results and Systematic Uncertainties}
\label{sec:results}
In this paper, we present a relative measurement; we extract the yields of 
both the signal mode $\overline{B} \to D^{(*)} \tau^+ \nu$ and the 
normalization mode $\overline{B} \to D^{(*)} \ell^+ \nu$ to deduce the four 
ratios,
\begin{eqnarray}
R(\overline{D}{}^0) & \equiv & 
{\cal B}\,(B^+ \to \overline{D}{}^0 \tau^+ \nu)/
{\cal B}\,(B^+ \to \overline{D}{}^0 \ell^+ \nu) \\
R(\overline{D}{}^{*0}) & \equiv & 
{\cal B}\,(B^+ \to \overline{D}{}^{*0} \tau^+ \nu)/
{\cal B}\,(B^+ \to \overline{D}{}^{*0} \ell^+ \nu) \\
R(D^-) & \equiv &
{\cal B}\,(B^0 \to D^- \tau^+ \nu)/
{\cal B}\,(B^0 \to D^- \ell^+ \nu) \\
R(D^{*-}) & \equiv & 
{\cal B}\,(B^0 \to D^{*-} \tau^+ \nu)/
{\cal B}\,(B^0 \to D^{*-} \ell^+ \nu).
\end{eqnarray}

The yields of the normalization modes are extracted as described in 
Section~\ref{sec:controlsample}.
For the signal modes, after finalizing the signal selection criteria and 
completing the studies in previous sections, we have opened the signal 
region, and performed the fits with the procedure described in 
Section~\ref{sec:fitting}.
Figures~\ref{fig:result_bc} and \ref{fig:result_bn} show 
the fit results for $B^+ \to D^{({*})} \tau \nu$ and 
$B^0 \to D^{({*})} \tau \nu$, respectively.
There are excesses in the signal region for all four decay modes.
Figure~\ref{fig:lh_curve} shows the signal likelihood curves,
while Table~\ref{tbl:results} summarizes the results.
The extracted yields (statistical significances) are 
$98.6^{+26.3}_{-25.0} (4.4)$, 
$99.8^{+22.2}_{-21.3} (5.2)$, 
$17.2^{+7.69}_{-6.88}  (2.8)$, and 
$25.0^{+7.17}_{-6.27} (5.9)$, for
$B \to \overline{D}{}^0 \tau^+ \nu$, $\overline{D}{}^{*0} \tau^+ \nu$,
$D^- \tau^+ \nu$ and $D^{*-} \tau^+ \nu$, respectively.
The efficiency $\epsilon$, shown in Table~\ref{tbl:results}, corresponds to 
the sum of the signal yields measured in $B \to D \tau \nu$ and 
$B \to D^* \tau \nu$ selections.
The ratio of ${\cal B}\,(B \to D^{(*)} \tau \nu)$ to 
${\cal B}\,(B \to D^{(*)} \ell \nu)$ are calculated as,
\begin{equation}
R(D^{(*)}) = 
\frac{N(D^{(*)} \tau \nu)}{N(D^{(*)} \ell \nu)}
\cdot 
\frac{2\epsilon(D^{(*)} \ell^+ \nu)}{\epsilon(D^{(*)} \tau^+ \nu)}
\cdot\frac{1}{{\cal B}\,(\tau\rightarrow\ell \nu \nu)}~~~~~.
\end{equation}
Note that the efficiency $\epsilon(D^{(*)} \ell \nu)$ is the average 
over electron and muon modes, while the yields are extracted for
the sum of the two modes.  

\begin{figure}[tbp]
\begin{center}
 \includegraphics[width=0.9\textwidth,keepaspectratio,clip]{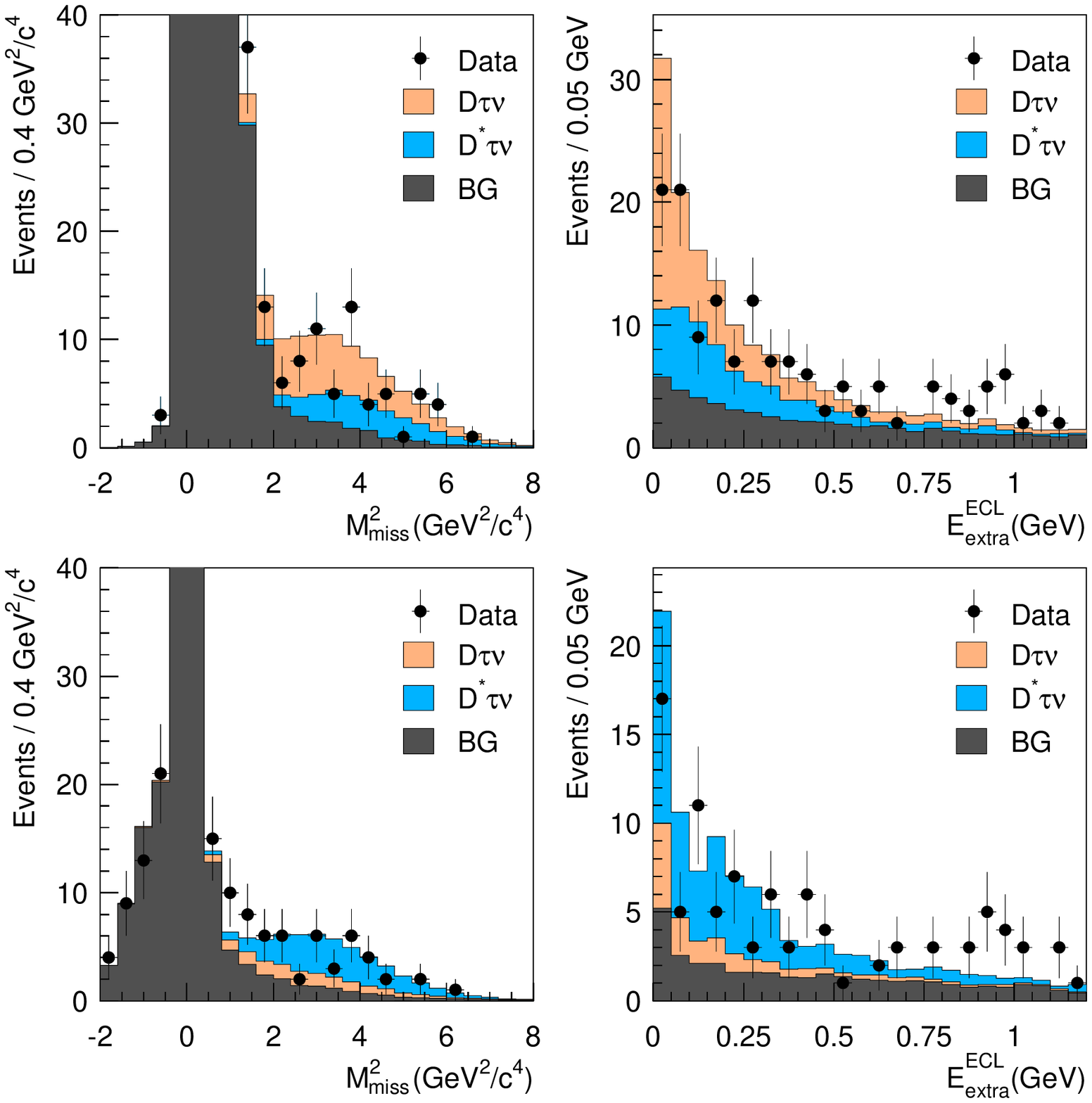}
\caption{
Fit results for $B^+ \to \overline{D}{}^0 \tau^+ \nu$ (top) and
$B^+ \to \overline{D}{}^{*0} \tau^+ \nu$ (bottom).
The $M_{\rm miss}^2$ (left) and $E_{\rm extra}^{\rm ECL}$ (right) distributions 
are shown with the signal selection cut on the other variable listed in 
Table~\ref{tbl:selection}.
}
\label{fig:result_bc}
\end{center}
\end{figure}

\begin{figure}[htbp]
\begin{center}
 \includegraphics[width=0.9\textwidth,keepaspectratio,clip]{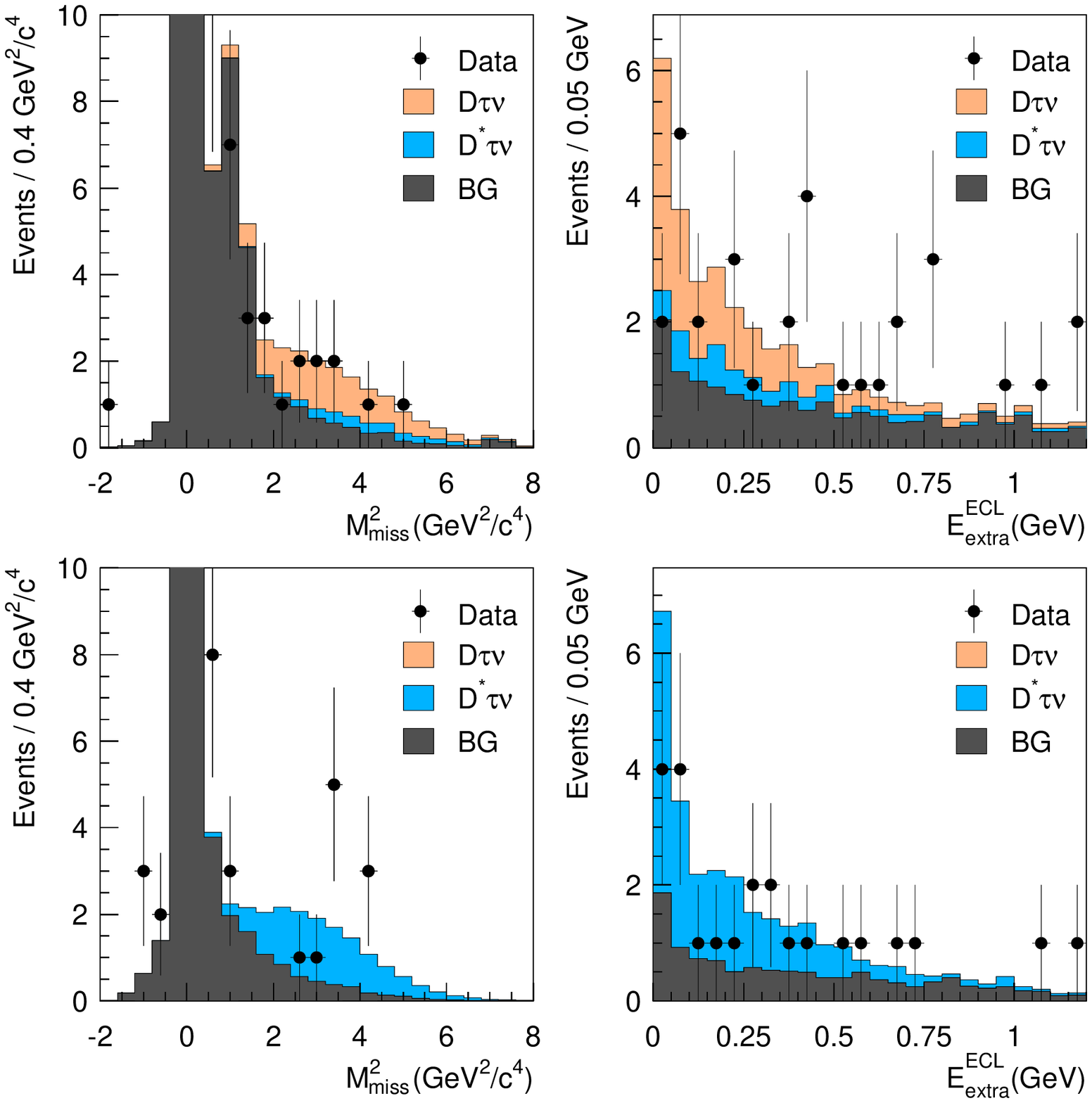}
\caption{
Fit results for 
$B^0 \to D^- \tau^+ \nu$ (top) and
$B^0 \to D^{*-} \tau^+ \nu$ (bottom).
The $M_{\rm miss}^2$ (left) and $E_{\rm extra}^{\rm ECL}$ (right) distributions 
are shown with the signal selection cut on the other variable listed in Table~\ref{tbl:selection}.
}
\label{fig:result_bn}
\end{center}
\end{figure}

\begin{figure}
\begin{center}
 \includegraphics[width=0.9\textwidth,keepaspectratio,clip]{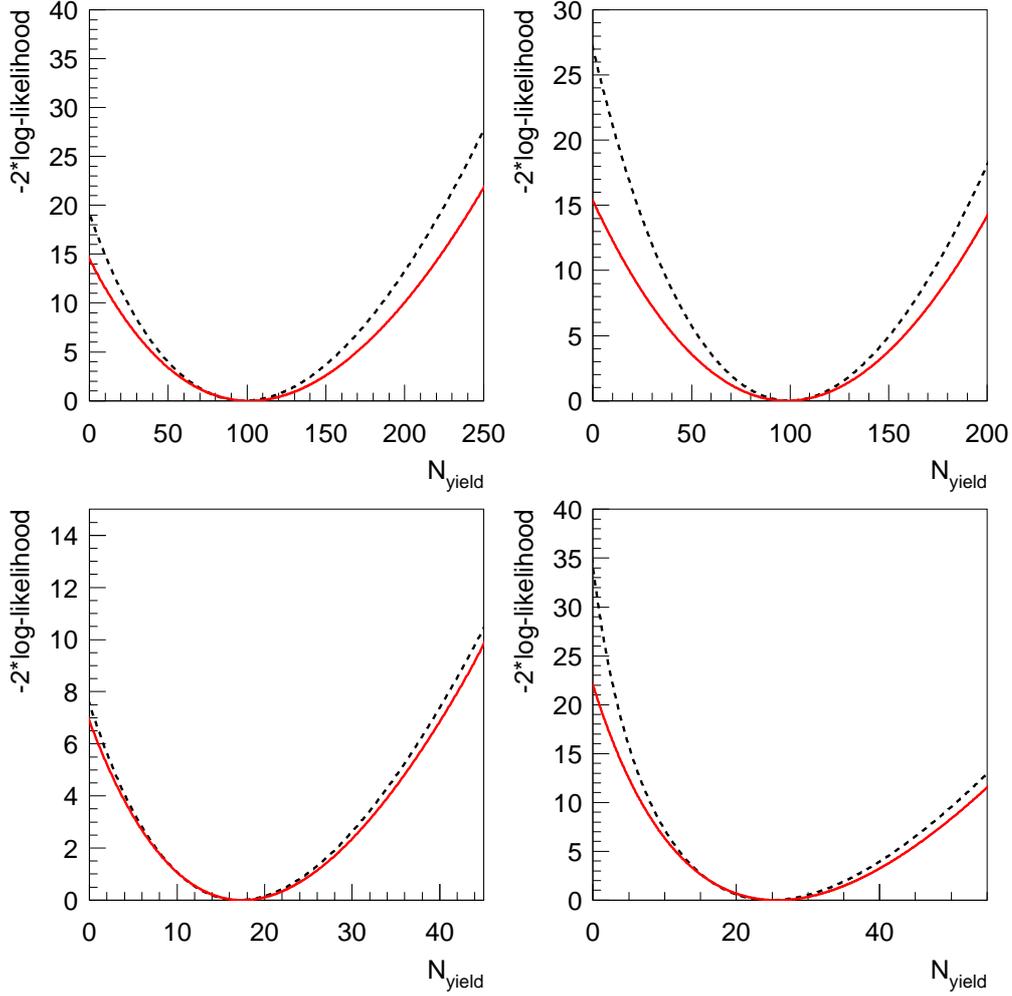}
\caption{
Likelihood curves as a function of signal yields for
$B^+ \to \overline{D}{}^0 \tau^+ \nu$ (top-left) and
$B^+ \to \overline{D}{}^{*0} \tau^+ \nu$ (top-right).
$B^0 \to D^- \tau^+ \nu$ (bottom-left) and
$B^0 \to D^{*-} \tau^+ \nu$ (bottom-right).
Red (black) curves show the likelihood 
with (without) the systematic uncertainty
}
\label{fig:lh_curve}
\end{center}
\end{figure}

\begin{table}[htbp]
 \begin{center}
  \begin{tabular}{lll}
   \hline\hline
   Quantity & 
   $\overline{D}{}^0 \tau^+ \nu$ &
   $\overline{D}{}^{*0} \tau^+ \nu$ \\
   \hline
   $N(\overline{D}{}^{(*)} \tau^+ \nu)$ & 98.6$^{+26.3}_{-25.0}$&
   99.8$^{+22.2}_{-21.3}$\\
   $\epsilon(\overline{D}{}^{(*)} \tau^+ \nu)$ [\%] &
   6.20 $\pm$ 0.08 & 3.86 $\pm$ 0.08   \\
   $R$[\%] & 70.2 $^{+18.9}_{-18.0}$ $^{+11.0}_{-9.1}$ & 
   46.8 $^{+10.6}_{-10.2}$  $^{+6.2}_{-7.2}$ \\
   $\Sigma (\Sigma_{stat})$  &3.8 (4.4) & 3.9 (5.2) \\
   ${\cal B}$ [\%]& 1.51 $^{+0.41}_{-0.39}$ $^{+0.24}_{-0.19}$  $\pm$ 0.15 &
   3.04 $^{+0.69}_{-0.66}$ $^{+0.40}_{-0.47}$ $\pm$ 0.22 \\
   \hline\hline

   Quantity & 
   $D^-    \tau^+ \nu$ &
   $D^{*-} \tau^+ \nu$ \\
   \hline

   $N(\overline{D}{}^{(*)} \tau^+ \nu)$ & 
   17.2$^{+7.7}_{-6.9}$ & 25.0$^{+7.2}_{-6.3}$ \\
   $\epsilon(\overline{D}{}^{(*)} \tau^+ \nu)$ [\%] &
    6.86 $\pm$ 0.09 & 2.09 $\pm$ 0.04   \\
   $R$[\%] &  47.6 $^{+21.6}_{-19.3}$  $^{+6.3}_{-5.4}$ &
   48.1 $^{+14.0}_{-12.3}$ $^{+5.8}_{-4.1}$ \\
   $\Sigma (\Sigma_{stat})$ & 2.6 (2.8) & 4.7 (5.9) \\
   ${\cal B}$[\%] & 
   1.01 $^{+0.46}_{-0.41}$  $^{+0.13}_{-0.11}$ $\pm$ 0.10 &
   2.56 $^{+0.75}_{-0.66}$  $^{+0.31}_{-0.22}$  $\pm$ 0.10 \\
   \hline\hline
  \end{tabular}
  \caption{Summary of the results; extracted yields from the fitting, $N$, 
the efficiency to detect the signal in either of 
$B \to D \tau (\ell) \nu$ and $B \to D^* \tau (\ell) \nu$ selections, $\epsilon$,  
the deduced ratio of ${\cal B}\,(B \to D^{(*)} \tau \nu)$ to 
${\cal B}\,(B \to D^{(*)} \ell \nu)$, $R$, 
significance of the signal with (without) systematic errors, $\Sigma (\Sigma_{stat})$,
deducued branching fraction, ${\cal B}$.}
  \label{tbl:results}
 \end{center}
\end{table}


Table~\ref{tbl:systematic} summarizes the systematic errors related to 
the ratio measurement, where reconstruction efficiency errors are 
largely cancel out. The following systematic errors are considered.

\begin{itemize}

\item {\bf $M_{\rm miss}^2$ shape:} 
The systematic error due to uncertainties in the $M_{\rm miss}^2$ shape is 
estimated by varying the PDF parameters.
The fitting procedure is repeated for each parameter variation, and
relative changes in the extracted yields are added in quadrature.
This method will give conservative estimates, as there are 
correlations in $M_{\rm miss}^2$ resolutions between decay modes.

\item {\bf $E_{\rm extra}^{\rm ECL}$ shape:} 
The systematic error due to uncertainties in the $E_{\rm extra}^{\rm ECL}$ 
shape is estimated by varying the content of each PDF histogram bin by its 
$\pm 1 \sigma$ statistical error.
The fitting procedure is repeated for each bin variation, and
relative changes in the extracted yields are added in quadrature.
 
\item {\bf $D^{**} \ell \nu$ branching fraction:} 
The systematic errors due to uncertainties in the $\overline{B} \to D^{**} 
\ell^+ \nu$ component is estimated by varying the branching fraction for 
each $D^{**}$ component by $\pm 1 \sigma$ based on the Belle results in 
\cite{Livent}.
The relative change in the extracted yields is assigned as the systematic 
error.
  
\item {\bf $D \leftrightarrow D^*$ cross-feed:}
In our nominal fitting procedure, the rates of the cross-feed between $D$ 
and $D^{*}$ decays are fixed to the values in the MC simulation, for
both the signal and normalization decays. 
The uncertainty is estimated by taking the relative change in the
extracted yield for the normalization decays, when the cross-feed
component is floated in the fit.  

\item {\bf $\tau \rightarrow \ell \nu \nu$ branching fraction:}
The systematic error due to uncertainties in the branching fraction of
$\tau$ decay modes is evaluated by changing the branching fractions
by the uncertainties in the PDG values~\cite{PDG2006}.

\end{itemize}
The total systematic error is the quadratic sum of all individual ones.

\begin{table}[htbp]
 \begin{center}
  \begin{tabular}{lllll}
   \hline\hline
   Source & 
   $\overline{D}{}^0 \tau^+ \nu$[\%] &
   $\overline{D}{}^{*0} \tau^+ \nu$[\%] &
   $D^-    \tau^+ \nu$[\%] &
   $D^{*-} \tau^+ \nu$[\%] \\
   \hline
   $M_{\rm miss}^2$ shape  & +9.10/-7.89 & +9.86/-10.7 & +6.39/-5.78 & +5.80/-6.12
   \\
   $E_{\rm extra}^{\rm ECL}$ shape 
   & +10.6/-7.58 & +7.01/-9.73 &
   +9.03/-7.27 & +9.84/-4.97 \\
   $D^{**} \ell \nu$ 
                & +0.35/-0.41 & +0.75/-0.02 & +4.50/-2.56 & +0.58/-0.28 \\
   $D \leftrightarrow D^*$ cross-feed & +7.05/-6.86 & +5.12/-5.34 & +5.77/-6.01
   & +3.48/-3.37 \\
   $\cal{B}(\tau \to \ell \nu \nu)$   & $\pm$0.3 & $\pm$0.3 & $\pm$0.3 & $\pm$0.3 \\ 
   \hline

   Total  & +15.7/-12.9 & +13.2/-15.4 & +13.3/-11.4 & +12.0/-8.58  \\ 

   \hline\hline
  \end{tabular}
  \caption{Summary of the systematic errors.}
  \label{tbl:systematic}
 \end{center}
\end{table}

With the systematic errors shown in Table~\ref{tbl:systematic}, the final 
results for the four ratios are,  
\begin{eqnarray}
R(\overline{D}{}^0) & = & 0.70~^{+0.19}_{-0.18}~^{+0.11}_{-0.09} \\
R(\overline{D}{}^{*0}) & = & 0.47~^{+0.11}_{-0.10}~^{+0.06}_{-0.07} \\
R(D^-) & = & 0.48~^{+0.22}_{-0.19}~^{+0.06}_{-0.05} \\
R(D^{*-}) & = & 0.48~^{+0.14}_{-0.12}~^{+0.06}_{-0.04} ~~~~~~,
\end{eqnarray}
where the first error is the statistical and the second error is the
systematic.
Including the systematic uncertainties for the yields convolved in the
likelihood\,(Figure~\ref{fig:lh_curve}), the significances of the excesses 
(in units of $sigma$) are found to be
$3.8, 3.9, 2.6$ and $4.7$ for 
$B \to \overline{D}{}^0 \tau^+ \nu$, $\overline{D}{}^{*0} \tau^+ \nu$,
$D^- \tau^+ \nu$ and $D^{*-} \tau^+ \nu$, respectively.

Using the branching fractions for the $B \to D^{(*))} \ell \nu$ normalization 
decays, reported in \cite{PDG2006}:
${\cal B}\,(B^+ \rightarrow D \ell \nu) = (2.15 \pm 0.22)$\%,
${\cal B}\,(B^+ \rightarrow D^* \ell \nu) = (6.5 \pm 0.5)$\%,
${\cal B}\,(B^0 \rightarrow D \ell \nu) = (2.12 \pm 0.20)$\%, and
${\cal B}\,(B^0 \rightarrow D^* \ell \nu) = (5.33 \pm 0.20)$\%,
we obtain the folowing branching fractions for $B \to D^{(*)} \tau \nu$ decays,
\begin{eqnarray}
 {\cal B}\,(B^+ \to \overline{D}{}^0 \tau^+ \nu) & = & 
  1.51 ~^{+0.41}_{-0.39}~^{+0.24}_{-0.19} \pm 0.15~[\%] \\ 
 {\cal B}\,(B^+ \to \overline{D}{}^{*0} \tau^+ \nu) & = & 
  3.04 ~^{+0.69}_{-0.66}~^{+0.40}_{-0.47} \pm 0.22~[\%] \\
 {\cal B}\,(B^0 \to D^- \tau^+ \nu) & = &
  1.01 ~^{+0.46}_{-0.41}  ~^{+0.13}_{-0.11} \pm 0.10~[\%] \\
 {\cal B}\,(B^0 \to D^{*-} \tau^+ \nu) & = &
  2.56~^{+0.75}_{-0.66} ~^{+0.31}_{-0.22} \pm 0.10~[\%] ~~~~~,
\end{eqnarray}
where the first error is statistical, the second is systematic, and the
third is due to the branching fraction error of the normalization mode.

\section{Summary}
Using 604.5 fb$^{-1}$ of data collected at the $\Upsilon(4S)$ resonance with 
the Belle detector at the KEKB collider, we have measured $B$ to $\tau$ semileptonic 
decays, by fully reconstructing hadronic decays of the accompanying $B$ meson.
We have extracted signals for the four decay modes, 
$B^+ \to \overline{D}{}^0 \tau^+ \nu$, 
$B^+ \to \overline{D}{}^{*0} \tau^+ \nu$,
$B^0 \to D^- \tau^+ \nu$, and 
$B^0 \to D^{*-} \tau^+ \nu$, and obtained the branching fractions 
listed in Section~\ref{sec:results}.
The obtained branching fractions are consistent within errors with the
earlier Belle result for $B^0 \to D^{*-} \tau^+ \nu$~\cite{Belle_Dsttaunu},
and BaBar results for the four signal modes \cite{BaBar_Dtaunu}.
Our results are slightly higher than the SM expectation, however more
luminosity is needed to clarify the deviation.

\vspace{1.0cm}
We thank the KEKB group for the excellent operation of the
accelerator, the KEK cryogenics group for the efficient
operation of the solenoid, and the KEK computer group and
the National Institute of Informatics for valuable computing
and SINET3 network support.  We acknowledge support from
the Ministry of Education, Culture, Sports, Science, and
Technology (MEXT) of Japan, the Japan Society for the 
Promotion of Science (JSPS), and the Tau-Lepton Physics 
Research Center of Nagoya University; 
the Australian Research Council and the Australian 
Department of Industry, Innovation, Science and Research;
the National Natural Science Foundation of China under
contract No.~10575109, 10775142, 10875115 and 10825524; 
the Department of Science and Technology of India; 
the BK21 program of the Ministry of Education of Korea, 
the CHEP SRC program and Basic Research program (grant 
No. R01-2008-000-10477-0) of the 
Korea Science and Engineering Foundation;
the Polish Ministry of Science and Higher Education;
the Ministry of Education and Science of the Russian
Federation and the Russian Federal Agency for Atomic Energy;
the Slovenian Research Agency;  the Swiss
National Science Foundation; the National Science Council
and the Ministry of Education of Taiwan; and the U.S.\
Department of Energy.
This work is supported by a Grant-in-Aid from MEXT for 
Science Research in a Priority Area (``New Development of 
Flavor Physics''), and from JSPS for Creative Scientific 
Research (``Evolution of Tau-lepton Physics'').


\end{document}